\documentclass[prd,aps,onecolumn,showpacs,superscriptaddress,floatfix,amsmath,footinbib,amssymb]{revtex4-1}

\usepackage[latin1]{inputenc}
\usepackage[english,spanish]{babel}
\usepackage{amsmath,amssymb}
\usepackage{graphicx}
\usepackage{hyperref}
\usepackage{orcidlink}

\newcommand{\eq}[1]{(\ref{#1})}
\newcommand{\ket}[1]{\vert #1 \rangle}

\renewcommand{\b}[1]{\mathbf{#1}}

\begin{document}

\title{Solid-state gravitational-wave detectors at GHz frequencies: the search for the primordial stochastic GW background and light primordial black hole binaries}

\author{Juan Garc\'ia-Bellido
\orcidlink{0000-0002-9370-8360}}
\email[]{juan.garciabellido@uam.es}
\affiliation{Instituto de F\'{i}sica Te\'orica UAM/CSIC, Universidad Aut\'onoma de Madrid, 28049 Madrid, Spain}

\begin{abstract}In modern cosmology the Big Bang is associated with the nonlinear transfer of energy from the inflaton to the other fields in the universe, at energies close to the GUT scale. This reheating after inflation is one of the strongest sources of gravitational waves (GW), producing a stochastic background (SGWB) with a non-thermal spectrum peaked at frequencies of order a few GHz. Detecting it is difficult: experiments based on the inverse Gertsenshtein effect in intense magnetic fields reach the MHz but not the GHz band, where the typical strain is around $10^{-30}$. The same window contains the coalescence of light primordial black hole (PBH) binaries, whose merger frequency $f\simeq4.4\ \mathrm{kHz}\,(M_\odot/M)$ falls in the MHz--GHz range for planetary to sub-planetary masses; since such objects are necessarily sub-solar, their detection would be strong evidence for PBHs as a component of the dark matter. We propose a solid-state detector at GHz frequencies that could integrate over months to years the GW continuously arriving from the Big Bang and search for light PBH binary coalescence. As a concrete realization we consider a modular array of $\sim10^3$ ultra-pure sapphire $(10\ \mathrm{cm})^3$ monocrystals forming a cubic-metre detector read out by cryogenic single-phonon sensors, whose segmentation provides thermal isolation, favourable counting statistics and coincidence-based background rejection. We also compare candidate materials, finding diamond superior per unit volume but limited by the unavailability of large single crystals. Finally, we contrast the two targets. The stationary background is a shot-noise-limited counting problem, best served by a narrow, resonance-enhanced, long-integration search; the loud transient chirp of a nearby merger is better caught by a fast, broad-band search with coincidence tagging. Because the phonon spectrum is continuous, a modular solid-state array can serve both, by staggering resonant cells across the band while running a broad-band subset for chirp tracking.
\end{abstract}

\maketitle

\section{Introduction}
The direct detection of gravitational waves (GW) by the LIGO and Virgo collaborations \cite{Abbott:2016} inaugurated a new era of observational astronomy, giving us access to the Universe through a channel entirely independent of electromagnetic radiation. Existing and planned observatories, from ground-based interferometers to the space-borne LISA mission and pulsar timing arrays, cover the frequency range from nHz to kHz, where a variety of astrophysical sources is expected. Above the kHz band the situation changes qualitatively: there are essentially no known astrophysical objects compact and dense enough to radiate appreciably at MHz--GHz frequencies. A detection of GW in the ultra-high-frequency (UHF) band would therefore most likely signal new physics beyond the Standard Model, associated either with exotic compact objects or with energetic processes in the very early Universe \cite{Aggarwal:2021,Aggarwal:2025}.

The cosmological motivation is strong. Causality relates the frequency of a primordially generated GW to the size of the comoving horizon at the time of production; GHz frequencies today correspond to the horizon at the highest energy scales accessible to particle physics, close to the scale of Grand Unification or string theory, i.e.\ to temperatures $T\gtrsim10^{10}\ \mathrm{GeV}$ \cite{Aggarwal:2025}. Among the strongest cosmological sources in this band is the non-linear stage of (p)reheating after inflation, during which the energy stored in the inflaton is transferred to the rest of the fields through parametric resonance and the collision of large field inhomogeneities \cite{GarciaBellido:2007}. This process generates a stochastic GW background (SGWB) with a non-thermal spectrum peaked, for typical inflaton potentials, at frequencies $f\sim10^{7}$--$10^{9}\ \mathrm{Hz}$ with a present-day energy density $\Omega_{\mathrm{GW}}\sim10^{-13}$--$10^{-11}$ \cite{Aggarwal:2025}. Other early-Universe processes, such as first-order phase transitions, cosmic strings and other topological defects, and the evaporation of light primordial black holes, contribute additional signals across the same band.

A second, complementary motivation comes from the late Universe. Although no ordinary astrophysical body is compact and dense enough to radiate above the kHz band, the coalescence of a binary of light primordial black holes (PBH) is an exception. As the binary inspirals and shrinks, the frequency of the emitted radiation grows and reaches, at the innermost stable circular orbit, $f_{\mathrm{ISCO}}\simeq4.4\ \mathrm{kHz}\,(M_\odot/M)$, where $M$ is the total mass, with the subsequent merger and ringdown radiating at higher frequencies. Binaries of planetary to sub-planetary mass therefore chirp across the MHz--GHz band: $M\sim10^{-3}\,M_\odot$ merges near $4\ \mathrm{MHz}$, $M\sim10^{-5}\,M_\odot$ near $0.4\ \mathrm{GHz}$ and $M\sim10^{-6}\,M_\odot$ near $4\ \mathrm{GHz}$. Since no object other than a black hole can be squeezed within its Schwarzschild radius at such small masses, a chirp detected in this band would be a clear signature of sub-solar, and hence primordial, black holes, objects that cannot form through ordinary stellar collapse and that may constitute a sizeable fraction of the dark matter in the asteroid- to planetary-mass window \cite{Aggarwal:2025}. A time-resolved detector sensitive to such transients would thus probe both the primordial background and light PBH dark matter.

The experimental challenge is severe. The predicted amplitudes translate into characteristic strains of order $h\sim10^{-30}$ or smaller at GHz frequencies, many orders of magnitude below the reach of present detectors and constrained only indirectly by the cosmological limit on the effective number of relativistic species. Over the past decade interest in UHF-GW detection has grown rapidly, and a broad landscape of detector concepts has emerged \cite{Aggarwal:2021,Aggarwal:2025}: reinterpretations of axion light-shining-through-wall and haloscope experiments through the inverse Gertsenshtein conversion of GW into photons in strong magnetic fields; resonant and heterodyne microwave cavities such as MAGO; levitated dielectric sensors; and resonant-mass acoustic devices. Of special relevance to the present work are the bulk acoustic wave (BAW) resonators pioneered by Goryachev and Tobar \cite{Goryachev:2014}, which use cryogenic quartz crystals with extremely high quality factors (up to $\sim10^{10}$) read out piezoelectrically with SQUIDs to search for GW in the MHz range, and the recent proposal to detect UHF-GW through single-phonon excitations in the same crystal targets developed for light dark matter searches \cite{Kahn:2024}. An earlier precursor of this line of thought is the proposal of \cite{Sabin:2014} to detect gravitational waves through the resonant creation of phonons in a Bose--Einstein condensate, where the passing wave modifies the effective (acoustic) metric on which the phonons propagate and excites quanta of the collective phononic field through a dynamical Casimir effect, with the wave amplitude estimated by relativistic quantum-metrology techniques. Although formulated for a quantum fluid rather than a crystalline solid, and aimed at the kHz band of astrophysical sources, it shares with the present work the idea of reading out at the single-quantum level the phonons created by a gravitational wave; its use of squeezed states also suggests a possible way to approach the shot-noise limit discussed below. These approaches share a common theme: the conversion of GW energy into the collective vibrational (phonon) degrees of freedom of a macroscopic quantum medium.

In this paper we develop, from first principles, the theory of a solid-state GW detector based on the coherent production of phonons in a crystalline solid. In contrast to the classic Weber-bar analysis, which treats the crystal as an elastic continuum characterized only by its macroscopic properties, we compute the GW-induced phonon production rate quantum mechanically, taking full account of the band structure of the lattice. Working in a general $d$-dimensional setting allows us to compare crystals of different dimensionality, from graphene (2D) to ordinary 3D crystals such as NaCl and sapphire (Al$_2$O$_3$). We show that Bragg's law renders the GW--matter interaction effectively one-dimensional, so that any extremum of a phonon band produces an integrable one-dimensional Van Hove singularity in the density of final states, which can resonantly enhance the transition rate. We then analyse the fundamental limitations --- thermal noise and, more importantly, the quantum shot noise associated with the discreteness of phonon production --- and estimate the performance of a realistic setup: a tower of $(10\ \mathrm{cm})^3$ sapphire monocrystals instrumented with cryogenic single-phonon sensors, in the spirit of the DAMA/ANAIS and CDMS dark-matter detectors \cite{2012RScI...83i1101L}.

The paper is organized as follows. We first review the Hamiltonian of a $d$-dimensional crystalline solid in terms of phonon degrees of freedom, and quantize it. We then derive the GW--crystal interaction and, via Fermi's golden rule, the phonon-production rate and the associated cross section. Next we present order-of-magnitude estimates of the signal and of the thermal background for graphene and NaCl, and show how a Van Hove singularity in the phonon density of states can resonantly enhance the rate. We then turn to the experimental prospects, proposing a concrete, modular realization --- an array of $\sim10^3$ ultra-pure sapphire $(10\ \mathrm{cm})^3$ monocrystals forming a cubic-metre detector read out by cryogenic single-phonon sensors --- and discussing the advantages of segmentation (thermal isolation, favourable counting statistics and coincidence-based background rejection) together with the choice of detector material, where we contrast sapphire with diamond. We next examine the complementary detection strategies demanded by the two target signals: a stationary stochastic background, whose shot-noise-limited, counting nature favours a narrow, resonance-enhanced, long-integration search, versus the loud transient chirp of a nearby PBH merger, which favours a fast, broad-band, coincidence-tagged one; we argue that a solid-state phonon detector is intrinsically broad-band and show how its band may be widened or tiled. We finally present our conclusions. A technical appendix collects the derivation of the GW--crystal interaction in a general transverse-traceless gauge.

\section{Crystal Hamiltonian}
In this section we review the Hamiltonian of a crystalline solid in terms of phonon degrees of freedom. The treatment here is standard and can be found in many texts \cite{Ashcroft,Kittel:ISSP}, although we include it here for completeness. The calculations are a $d$ dimensional version of the linear chain considered in \cite{grischuk}; by considering the general case of a $d$-dimensional crystal we are able to compare systems of different dimensionality, such as graphene (2D) and ordinary (3D) crystals. However, note that, even if we allow the dimension of the lattice to vary, we always work in a 4-dimensional spacetime.

We start by setting up some notation. We will consider a monochromatic gravitational wave  $h_{\mu\nu} =h\epsilon_{\mu\nu}e^{i(\b{k}_g\b{x}-\omega_g t)}$ in transverse traceless gauge,  scattering off a crystal of $N$ unit cells. These cells will be spanned by a set of basis vectors $\{\b{a}_n\}_{n=1\ldots d}$. The $i$th component of $\b{a}_n$ will be denoted $a_n^i$.  A general element $\b{R}$ of the lattice therefore can be expanded as $\b{R}=\sum_n e_n\b{a}_n,\quad n\in\mathbb{N}$. Similarly, the basis of the dual lattice to $\{\b{a}_n\}$ will be denoted $\{\b{b}_n\}_{n=1\ldots d}$. Crystal momenta live in the first Brillouin zone of the dual lattice \cite{Ashcroft} and may therefore be written as $\b{k}=\sum_nk_n\b{b}_n$.

Each cell will be labeled by a lattice vector $\b{R}$ and will contain $\ell$ atoms, possibly of different masses $m_l$ ($l=1\ldots \ell$), whose position vectors with respect to the origin of the cell are denoted by $\b{c}_l$. These atoms will be perturbed from their equilibrium positions due to any external perturbation, such as the gravitational wave. The main degree of freedom we are concerned with is the displacement of the $l$-th atom from its equilibrium position, and will be labeled by a vector $\b{u}_l(\b{R})$. The crystal will be subject to periodic boundary conditions,  as usual in solid state physics (see e.g. \cite{Ashcroft,Kittel:ISSP}). That is, we will require that $\b{u}_l(\b{R})=\b{u}_l(\b{R})+N_n\b{a_n}$ for some prescribed set of integers $N_n$, which specify the length of the crystal along each basis direction. In the analysis of Weber bars, it is customary \cite{PhysRev.117.306,grischuk} to consider free boundary conditions at the end of the crystal. This is necessary if the study is centered on frequencies near the fundamental mode of the bar. This is not the case here, where we will consider very high frequencies compared to the fundamental mode. As a result, phonon wavepackets are very well localized within the crystal and periodic boundary conditions can be safely imposed.

The interaction between the atoms in the solid can be cast in the form \cite{Ashcroft}
\begin{align}\sum_{\b{R},l,\b{R}',l'} u^j_{l'}(\b{R}') D^{ll'}_{ij}(\b{R}+\b{c}_l-\b{R}'-\b{c}_{l'})u^i_l(\b{R}).\end{align}
The matrix $D_{ij}^{ll'}$ is symmetric in both pairs of indices and it is related to the second-order expansion of the full interaction between atoms of type $l,l'$. We note that $D_{ij}(0)=0$, as such a term would violate invariance under lattice translations. The total Hamiltonian is the sum of the interaction plus the kinetic term for each atom,
\begin{align}\label{ham}H=\sum_{\b{R},l}\frac{\vert \b{p}_{l}(\b{R})\vert^2}{2m_l}+\frac{1}{2}\sum_{\b{R}',l'} u^j_{l'}(\b{R}') D^{ll'}_{ij}(\b{R}+\b{c}_l-\b{R}'-\b{c}_l')u^i_l(\b{R}).\end{align}
As the Hamiltonian is a quadratic form in coordinates and momenta, it is trivial to diagonalize it . To achieve this we first define the new set of variables $\b{u}'_l=\sqrt{m_l}\b{u}_l$ . This implies $\b{p}'_l=\sqrt{m_l}\b{p}_l$, which has the effect of turning the kinetic term in \eq{ham} into a multiple of the identity. Hence, now we only have to diagonalize the new interaction term finding the appropriate set of eigenfunctions. This is most easily done using the equations of motion, which are, using the symmetry properties of $D_{ij}$,
\begin{align}m_l\ddot{u}^i_l(\b{R})=-\sum_{\b{R}',l'}D^{ll'}_{ij}(\b{R}+\b{c}_l-\b{R}'-\b{c}_l')u^j_{l'}(\b{R}').\end{align}
If we introduce an ansatz of the form
\begin{align}u^i_l(\b{R})=\epsilon^i_{l,m}(\b{k})e^{i[\b{k}\b{R}-\omega_m(\b{k})t]}\end{align}
where, to comply with the periodic boundary conditions, $\b{k}$ is of the form
\begin{align}\b{k}=\sum_n \frac{r_n}{N_n}\b{b}_n,\quad r_n\in \mathbb{N},\label{peri}\end{align}
we arrive at
\begin{align}m_l\omega_m(\b{k})^2\epsilon^i_{l,m}(\b{k})=\left[\sum_{\b{R}',l'}D^{ll'}_{ij}(\b{R}+\b{c}_l-\b{R}'-\b{c}_l')e^{i\b{k}\b{R}'}\right]\epsilon^j_{l',m}(\b{k}).\end{align}
This is a matrix eigenvalue problem for the vectors $\epsilon^i_{l,m}(\b{k})$ in the $3\ell$-dimensional space spanned by the indices $i$ and $l$, where $m\in\{1\ldots 3\ell\}$ labels the eigenvalues. The hermitian matrix to be diagonalized is, in terms of primed variables,
\begin{align}D_{ij}^{ll'}(\b{k})=\frac{1}{\sqrt{m_lm_l'}}\sum_{\b{R}'}D^{ll'}_{ij}(\b{R}+\b{c}_l-\b{R}'-\b{c}_l')e^{i\b{k}\b{R}'}.\end{align}
Polarization eigenvectors $\epsilon^i_{l,m}(\b{k})$ can therefore be chosen so that they obey an orthogonality condition of the form
\begin{align}\sum_{j,l}m_l\epsilon^j_{l,m}(\b{k})\epsilon^j_{l,m'}(\b{k})^*=\bar{m}\delta_{m,m'},\quad \bar{m}=\frac{1}{\ell}\sum_lm_l.\end{align}

With these eigenvectors, the general real solution to the Hamiltonian \eq{ham} may be written as
\begin{align}\label{fi}u_l^i(\b{R})=\sum_{\b{k},m}\left( \alpha_m(\b{k}) \epsilon^i_{l,m}(\b{k})e^{i[\b{k}\b{R}-\omega_m(\b{k})t]}+\alpha^*_m(\b{k}) \epsilon^i_{l,m}(\b{k})^*e^{i[-\b{k}\b{R}+\omega_m(\b{k})t]}\right).\end{align}

We now substitute \eq{ham} in terms of \eq{fi}. The kinetic term reads
\begin{align}T&=-\frac{1}{2}\sum_{l,\b{R}}\sum_{\b{k},m,\b{k}',m'}\omega_m(\b{k})^2\left[\alpha_m(\b{k})\alpha_{m'}(\b{k}') \epsilon^i_{l,m}(\b{k})\epsilon^i_{l,m'}(\b{k}')e^{i[(\b{k}+\b{k}')\b{R}-(\omega_m(\b{k})+\omega_{m'}(\b{k}'))t]}\right.\nonumber\\&\left.+\alpha^*_m(\b{k})\alpha^*_{m'}(\b{k}') \epsilon^i_{l,m}(\b{k})^*\epsilon^i_{l,m'}(\b{k}')^*e^{-i[(\b{k}+\b{k}')\b{R}-(\omega_m(\b{k})+\omega_{m'}(\b{k}'))t]}\right.\nonumber\\&\left.-\alpha_m(\b{k})\alpha^*_{m'}(\b{k}') \epsilon^i_{l,m}(\b{k})\epsilon^i_{l,m'}(\b{k}')^*e^{i[(\b{k}-\b{k}')\b{R}-(\omega_m(\b{k})-\omega_{m'}(\b{k}'))t]}\right.\nonumber\\&\left.-\alpha^*_m(\b{k})\alpha_{m'}(\b{k}') \epsilon^i_{l,m}(\b{k})^*\epsilon^i_{l,m'}(\b{k}')e^{i[(-\b{k}+\b{k}')\b{R}-(-\omega_m(\b{k})+\omega_{m'}(\b{k}'))t]}\right]\nonumber\\&=N\bar{m}\sum_{\b{k},m}\omega_m(\b{k})^2\left[\vert\alpha_m(\b{k})\vert^2-2\cos(\omega_m(\b{k})t)\alpha_m(\b{k})\alpha_m(-\b{k})\right].\end{align}
To reach this expression, we have used the fact that $\epsilon^i_{l,m}(\b{k})^*=\epsilon^i_{l,m}(-\b{k})$, for real displacements $\b{u}_l^i(\b{R})$. The interaction term gives an identical but opposite contribution to the time-dependent term, and as a result the Hamiltonian is
\begin{align}H=2N\sum_{\b{k},m}\bar{m}\omega_m(\b{k})^2\vert\alpha_m(\b{k})\vert^2.\end{align}
We are now in a position where quantization of the Hamiltonian is straightforward. We promote the coefficients $\alpha^*_m(\b{k})$ to creation operators according to the rule
\begin{align}\alpha^*_m(\b{k})\rightarrow \alpha^\dagger_m(\b{k})=\sqrt{\frac{\hbar}{2N\bar{m}\omega_m(\b{k})}}a^\dagger_m(\b{k})\end{align}
so that the Hamiltonian takes the expected form in terms of phonons,
\begin{align}H=\sum_{\b{k},m}\hbar\omega_m(\b{k})a^\dagger_m(\b{k})a_m(\b{k}),\end{align}
and the atomic displacement in the Schr\"odinger picture takes the form
\begin{align}\label{qfi}u_{l,m}^i(\b{R})=\sum_{\b{k},m}\sqrt{\frac{\hbar}{2N\bar{m}\omega_m(\b{k})}}\left( a^\dagger_m(\b{k}) \epsilon^i_{l,m}(\b{k})e^{i\b{k}\b{R}}+a_m(\b{k}) \epsilon^i_{l,m}(\b{k})^*e^{-i\b{k}\b{R}}\right).\end{align}

\section{Interaction with the gravitational wave}
The interaction term between the gravitational wave and the solid described above will be, on very general terms, of the form $-\frac12\int h_{ab}T^{ab}$, where $T^{ab}$ is the stress-energy tensor of the solid. However, as done in \cite{grischuk,Branchina:2004wk}, it is more straightforward to proceed in a different way, by taking into account the specific tidal forces acting on each atom and obtaining a potential for them. This cannot be done unless a coordinate system can be specified. If the GW wavelength is much longer than the crystal dimensions, we can always pick a reference frame comoving with the center of mass of the crystal. In this frame, the interaction between the atoms and the GW is just an oscillatory motion generated by the geodesic deviation equation. In the linear approximation, this is just  a force 
\begin{align}\frac{m}{2} \ddot{h}_{ij}x^j\end{align} acting on a particle sitting at the position $\b{x}$ as measured from the center of mass. This may be described by a potential term
\begin{align}-\frac{m}{4} \ddot{h}_{ij}x^jx^i\label{pif}\end{align}
For our crystal, the position of a generic atom in the lattice is $x^i= R^i+c_l^i+u^i_l(\b{R})$. Since $R^i+c_l^i\gg u^i_l(\b{R})$, the  perturbation to the Hamiltonian is, to a linear approximation for the displacement,
\begin{align}-\frac{m_l}{2} \ddot{h}_{ij}[R+c_l]^ju_l^i(\b{R}).\end{align}
Adding up the contributions of all the atoms and using \eq{qfi}, the total perturbation induced by a GW with a wavelength much longer than the crystal dimensions is
\begin{align}\label{P}\Delta V=-\frac{\ddot{h}_{ij}}{2}\sum_{\b{R},l,\b{k},m}m_l\sqrt{\frac{\hbar}{2N\bar{m}\omega_m(\b{k})}}[R+c_l]^j\left( a^\dagger_m(\b{k}) \epsilon^i_{l,m}(\b{k})e^{i\b{k}\b{R}}+a_m(\b{k}) (\epsilon^*)^i_{l,m}(\b{k})e^{-i\b{k}\b{R}}\right).\end{align}
The treatment so far has been concerned only with gravitational wavelengths much longer than the size of the crystal. However, for GW backgrounds produced at the end of inflation, if the results of the BICEP2 collaboration hold \cite{Ade:2014xna} we expect wavelengths of order $\sim 1\ cm$, which cannot be taken as much longer than a macroscopic crystal of tens of cm as we intend. As developed in detail in appendix \ref{appendixgauge}, it turns out that, as long as the gravitational wavelength is much larger than the lattice spacing, \eq{P} can still be taken as the correct perturbation for the system under consideration, with the only proviso that $e^{-i\b{k}\b{R}}$ must be replaced by $e^{i[\b{k}-\b{k}_g]\b{R}}=e^{i\b{k}'\b{R}}$. In other words, the gravitational wave has a nonvanishing momentum transfer to the crystal. From now on, we will consider this form for the perturbation.

The perturbation may be rewritten in a simpler form by performing first the summation over $\b{R}$. The $\b{R}$-dependence in the positive frequency part of \eq{P} is
\begin{align}\sum_{\b{R}}[\b{R}+\b{c}_l]^je^{i\b{k}'\b{R}}=\sum_{\b{R}}\b{R}e^{i\b{k}'\b{R}}.\end{align}
The $\b{c}_l$-dependent term disappears because for the $\b{k}$'s allowed by \eq{peri}, the sum adds up to $\delta_{\b{k}',0}$ and since we are assuming that the GW wavelength is much larger than the atomic dimensions, this basically states that $\b{k}=0$. This corresponds to center-of-mass motions of the whole crystal, which will not take place in the center-of-mass frame, and so the term may be neglected. The resulting sum can be evaluated if we express all the vectors involved in terms of a lattice basis and a dual lattice basis,
\begin{align}\sum_{\b{R}}\b{R}e^{i\b{k}'\b{R}}&=\sum_{e_1,e_2\ldots e_n}(e_1\b{a}_1+\ldots e_n\b{a}_n) e^{2\pi i e_1 k_1'}\cdot\ldots e^{2\pi i e_n k_n'}=\sum_n\left[ \left(\sum_{e_n}e_n\b{a}_n e^{2\pi i k_n'e_n }\right)\prod_{n'\neq n}\left(\sum_{e_{n'}}e^{2\pi i e_{n'} k_{n'}}\right)\right]\nonumber\\&=\frac{1}{2\pi i}\sum_n\left[  \b{a}_n\frac{d}{dk'_{n}}\left(\sum_{e_n}e^{2\pi i k_n'e_n }\right)\prod_{n'\neq n}\left(\sum_{e_{n'}}e^{2\pi i e_{n'} k_{n'}}\right)\right].\label{krtrans}\end{align}
Here we have turned the sum over $n'$ into a product and then evaluated each of the factors separately, except for one. The sums are evaluated as follows:
\begin{align}\sum_{e_{n'}=1}^{N_{n'}}e^{2\pi i e_{n'} k_{n'}}=e^{2\pi i k_{n'}} \frac{1-e^{2\pi i k_{n'} N_{n'}}}{1-e^{2\pi i k_{n'}}}=0\quad  \text{ for $k_{n'}\neq0$},\end{align}
where in the last equality we have used the fact that, by virtue of the boundary conditions, $k_{n'} N_{n'}$ is always an integer. For $k_{n'}=0$, the sum is just $N_{n'}$ and so in general it evaluates to the discrete Kronecker delta $N_{n'}\delta_{k_{n'},0}$. 

The only term left is therefore
\begin{align}\frac{1}{2\pi i N_n}\frac{d}{d k'_n}\left(\sum_{e_n}e^{2\pi ik'_{n}e_n}\right)=\frac{e^{2\pi i k'_n}}{e^{2\pi i k'_n}-1},\end{align}
where again we have used the fact  that $k'_nN_n$ is an integer, and also that the sum goes from $1$ to $N_n$. Finally, the perturbation may be written as 
\begin{align}\label{Pu}\Delta V=-\frac{\ddot{h}_{ij}}{2}\sum_{l,\b{k},m}m_l\sqrt{\frac{N\hbar}{2\bar{m}\omega_m(\b{k})}} a^\dagger_m(\b{k}) \epsilon^i_{l,m}(\b{k})\left[\sum_n a_n^j \left(\frac{e^{2\pi i k'_n}}{e^{2\pi i k'_n}-1}\right)\delta^{\perp}_{\b{k}',\b{b}_n}\right] + (\text{H.c.}).\end{align}
Here we have introduced a `transverse' delta $\delta^\perp_{\b{k},\b{q}}$ which takes the value 1 if both arguments are parallel vectors and 0 otherwise, and we remind the reader that $N\equiv N_1\ldots N_n$. An interesting result is that the perturbation \eq{Pu} can only excite, to first order in perturbation theory, phonon modes such that $\b{k}=\b{k}_g+\b{b}_n$, i.e. the phonon momentum must be related to the GW momentum by a reciprocal lattice vector. This is nothing else than Bragg's law and reflects the periodic nature of the lattice.

Using \eq{Pu} we can compute, using Fermi's golden rule, the amount of phonons per second created by the scattering of gravitational wave off the crystal. For this we need to take a continuum limit, $N\rightarrow\infty$, since only in this case the rule can be applied exactly. The total transition rate is \cite{Fermi:1950:NPCa,Bal}
\begin{align}\Gamma &=\frac{1}{T}\sum_{\b{k},m}\frac{1}{\hbar^2}\left\vert \int_0^T\langle k\vert P \vert 0\rangle dt\right\vert^2\nonumber\\&=\frac{1}{T}\sum_{\b{k},m}\left(\frac{N h^2}{2\hbar\bar{m}\omega_m(\b{k})}\right)\left\vert\frac{\epsilon_{ij}}{2}\sum_{l,m}m_l \epsilon^i_{l}(\b{k})\left[\sum_n a_n^j \left(\frac{e^{2\pi i k'_n}}{e^{2\pi i k'_n}-1}\right)\delta^{\perp}_{\b{k}',\b{b}_n}\right]\int_0^T e^{i(\omega_m(\b{k})-\omega_g)t}dt\right\vert^2\nonumber\\&\approx\sum_{\b{k},m,n}\left(\frac{\pi N h^2\omega_g^4\bar{m}\vert a_n\vert^2}{4\hbar\omega_m(\b{k})}\right)\left\vert\epsilon_{ij}\sum_{l}\frac{m_l}{\bar{m}} \epsilon^i_{l,m}(\b{k})\frac{a_n^j}{\vert a_n\vert} \left(\frac{1}{e^{2\pi i k'_n}-1}\right)\delta^{\perp}_{\b{k}',\b{b}_n}\right\vert^2\delta(\omega_m(\b{k})-\omega_g)\nonumber\\&\approx\sum_{m,n}\int N_n dk_n\left(\frac{\pi N h^2\omega_g^4\bar{m}\vert a_n\vert^2}{4\hbar\omega_m(k_n\b{b}_n)}\right)\left\vert\epsilon_{ij}\sum_{l}\frac{m_l}{\bar{m}} \epsilon^i_{l,m}(k_n\b{b}_n)\frac{a_n^j}{\vert a_n\vert} \left(\frac{1}{e^{2\pi i k'_n}-1}\right)\right\vert^2\delta(\omega_m(k_n\b{b}_n)-\omega_g).\end{align}
The first line is just the definition of the transition rate, from vacuum to one-particle state. In the second line we have computed the matrix element between the final and initial states. In the third we have used the fact that the transverse deltas guarantee the absence of mixed terms, which allows us to take the $n'$ sum out of the matrix element. Finally, following a standard procedure, we have evaluated the time integrals to $2\pi T\delta(\omega_m(\b{k})-\omega_g)$.We also have the average mass $\bar{m}=\ell^{-1}\sum_l m_l$ in order to make the matrix element dimensionless.

The only thing left is the $k_n$ integral. Using that 
\begin{align}\delta(\omega_m(\b{k})-\omega_g)=\sum_{p}\frac{1}{\vert \nabla \omega_m(\b{k})\cdot \b{b_n}\vert}\delta(k_n-k_{n,m,p}),\end{align}
where $k_{n,m,p}$ is a solution of the equation $\omega_m(k_n\b{b}_n)=\omega_g$, with $p$ an index labelling the possible solutions to the equation (the different bands). Since all lattices have inversion symmetry, if $k_{n,m,p}$ is a solution then so will be $-k_{n,m,p}$. We define $\b{k}_{n,m,p}=k_{n,m,p}\b{b}_n$, and $L_n=\vert a_n\vert N_n$ is the size of the crystal along the $n$-th direction. The integral can be readily evaluated, and we have
\begin{align}\label{tr}\Gamma &=\sum_{m,n,p}\left(\frac{\pi N h^2\omega_g^3\bar{m}L_n\vert a_n\vert}{8\hbar}\right)\left\vert\epsilon_{ij}\sum_{l}\frac{m_l}{\bar{m}} \epsilon^i_{l,m}(\b{k}_{n,m,p})\frac{a_n^j}{\vert a_n\vert}\right\vert^2\frac{1}{2\left[1-\cos(2\pi k'_n)\right]}\frac{1}{\vert \nabla \omega_m(\b{k}_{n,m,p})\cdot \b{b_n}\vert}.\end{align}
This is the result we were looking for, the number of phonons produced by the GW per unit time as it traverses the crystal. A similar expression for the one-dimensional case can be found in \cite{grischuk}.

To estimate the order of magnitude of \eq{tr}, we will make some simplifying assumptions: The tensor contractions involving polarization vectors will be assumed to be of order one, and take a square/cubic lattice with $\vert\b{a}_n\vert=a$, as well as $L_n=L$. We will assume that the dispersion relation is such that there is only one relevant branch which furthermore is acoustic, so that $\omega_g=c_s\b{k}$. Notice that this assumption does not hold if there are several branches with nearly degenerate $c_s$; however, for order of magnitude calculations, the one-branch assumption is reasonable. Finally, we will also assume that $k_n'\approx a\omega_g/(2\pi c_s)$ is very small so that the cosine in \eq{tr} may be expanded to second order. Under these assumptions, we have
\begin{align}\Gamma\sim\frac18\frac{\bar{m}ac_s}{\hbar}\left(\frac{L}{a}\right)^{d+1}h^2\omega_g,\label{gapprox}\end{align}
or, in terms of energy per unit time (power),
\begin{align}P\sim\frac18\bar{m}ac_s\left(\frac{L}{a}\right)^{d+1}h^2\omega^2_g.\label{edep}\end{align}
This expression, as well as  \eq{tr}, is a tree-level result (i.e. it is $\hbar$-independent), and could also be obtained through a classical analysis\cite{grischuk}. The advantage of doing a full quantum-mechanical calculation is that it takes into account the full band structure of the solid. Since the energy flux of a monochromatic plane GW is
\begin{align}I=\frac{c^3}{16\pi G} h^2\omega_g^2,\end{align}
the phonon production cross section of the detector is
\begin{align}\sigma=\frac{P}{I}=\frac{2\pi G}{c^3}\bar{m}ac_s\left(\frac{L}{a}\right)^{d+1}.\end{align}

\begin{figure}[!hbtp]
\centering
\includegraphics[width=0.65\textwidth]{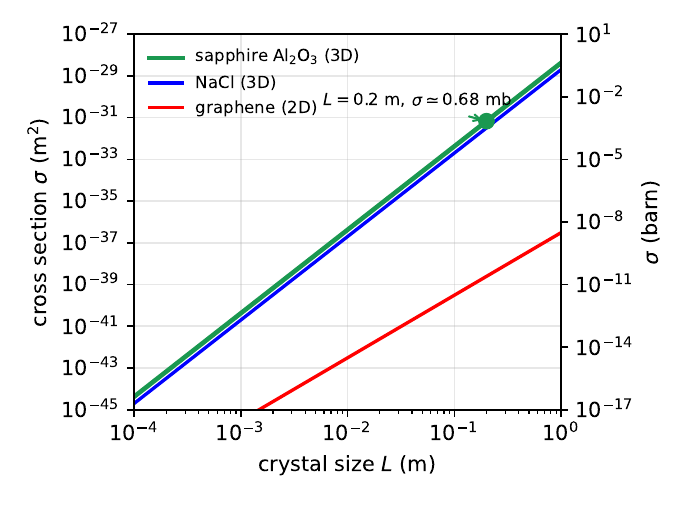}
\caption{Phonon production cross section $\sigma$ (in $\mathrm{m}^2$, right axis in barn) as a function of crystal length $L$ for sapphire Al$_2$O$_3$ (green), NaCl (blue) and graphene (red). The coherent scaling $\sigma\propto(L/a)^{d+1}$ makes the cross section grow rapidly with $L$: at $L\sim 1\ m$ it reaches $\sigma\sim2\cdot10^{-29}\ \mathrm{m}^2$ ($\sim0.2\ \mathrm{barn}$) for NaCl and $\sigma\sim3\cdot10^{-37}\ \mathrm{m}^2$ for graphene. The green dot marks the sapphire operating point $L=0.2$ m used in the experimental proposal, $\sigma\simeq0.68$ mb. For macroscopic crystals the higher dimensionality of the 3D crystals wins over the larger sound speed and smaller lattice spacing of graphene.}
\label{f1}
\end{figure}

Figure \ref{f1} shows the cross section for a wide range of detector crystal length, for three examples: sapphire (Al$_2$O$_3$), common salt (NaCl) and graphene. Even though the much larger sound speed of graphene ($c_s\approx2\cdot10^4\ m/s$) gives it an edge over salt at low $L$, the different dimension of both crystals eventually makes the 3D crystals the better choice for macroscopic detectors. Among the 3D options, sapphire is particularly attractive as an experimental material: it can be grown as large, ultra-pure monocrystals, has a high sound speed $c_s\approx10^4\ m/s$, lattice spacing $a\approx5\cdot10^{-10}\ m$ and average atomic mass $\bar{m}\approx3.4\cdot10^{-26}\ kg$ (the mean of the Al and O masses in Al$_2$O$_3$). For a sapphire crystal of size $L=0.2$ m these values give a coherent cross section $\sigma\simeq0.68\ mb$, marked as the operating point in Fig.~\ref{f1}.

\section{Some estimates}
We will now use the order-of-magnitude estimate \eq{edep} to study the feasibility of solid state detectors tuned to relic GW backgrounds. The fact that we did not specify a particular value of $d$ allows us to compare systems with different lattice dimensionality, such as graphene (2D) and ordinary crystals (3D).

For a given crystal, the signal strength is the result of the comparison of two competing factors: The GW-induced phonons and the thermal background. To estimate this background, consider a $d$-dimensional system of massless free particles with $p$ degrees of freedom. This describes a phonon system in a perfect crystal; interactions which will generate scattering between the different branches have been neglected. The energy density per unit wavelength is
\begin{align}n_\lambda= S(d)\cdot 2\pi\hbar pc_s\frac{\lambda^{-(d+2)}}{e^{\frac{2\pi\hbar c_s}{\lambda k_BT}}-1}.\end{align}
where $S(d)$ is the surface of the unit sphere in $d$ dimensions. From the energy density we can compute the Planck distribution by multiplying by $c_s$ and dividing by $S(d)$ to obtain the energy flux per unit solid angle. In this way we recover Planck's law as 
\begin{align}B_\lambda= 2\pi\hbar pc_s^2\frac{\lambda^{-(d+2)}}{e^{\frac{2\pi \hbar c_s}{\lambda k_BT}}-1}.\end{align}
To obtain the total background we should integrate the above expression in the relevant range of wavelengths and multiply by the detector area (which we will take to scale with the boundary of the system, as $L^{d-1}$). An estimate can be obtained by integrating over the frequency range of the gravitational wave. Under the assumption that the incoming radiation is almost monochromatic, we may evaluate the background over a window of wavelength width $\sigma_\lambda$, which would correspond to the spectral resolution of the detector. We obtain
\begin{align}\frac{\Gamma_T}{L^{d-1}}=S(d)\int_0^{\frac{\pi}{2}}\sin^{d-2}\theta d\theta \int_0^\infty d\lambda \left(\frac{1}{2\sigma_\lambda}\right)\delta (\lambda-\lambda_g) B_\lambda=\frac{1}{\sigma_\lambda}\pi\hbar pc_s^2S(d)\frac{\sqrt{\pi}\Gamma\left(\frac{d-1}{2}\right)}{\Gamma(d/2)}\frac{\lambda_0^{-(d+2)}}{e^{\frac{2\pi \hbar c_s}{\lambda k_BT}}-1}\equiv f(T).\end{align}
The thermal transition rate will be $\Gamma_T=L^{d-1}f(T)$. With this expression we are finally able to compute the interesting quantity from the experimental point of view, the signal-to-noise ratio, assuming most of the noise comes from thermal fluctuations, as
\begin{align}R&=\frac{\Gamma}{\Gamma_T}=\frac18\bar{m}ac_s\frac{L^2}{a^{d+1}f(T)}h^2\omega^2_g.\end{align}

This estimate allows us, for a given crystal and signal to noise ratio, to estimate the maximum temperature compatible with the specified value of $R$. We now proceed to compare this temperature for two crystals of different dimensionality, graphene and sodium chloride. These computations will assume, for the sake of simplicity, $\sigma_\lambda\sim\lambda_g/10$; the results depend only weakly (logarithmically) on this assumption.

For graphene, we have $d=2$, $\bar{m}=12$ a.m.u, $a\sim 0.142\ nm$ and $c_s\sim2\cdot10^4 m/s$ \cite{Cooper:2011}, so we have, for a $L=1\ m$ crystal, a signal-to-noise ratio $R\sim1$ and a GW amplitude $h\sim10^{-30}$, a maximum temperature of $T_\text{max}^\text{graphene}\approx 2\cdot10^{-2}\ K$.

For sodium chloride, $d=3$, $\bar{m}=29$ a.m.u, $a\sim 0.56\ nm$ and $c_s\sim4.7\cdot10^3 m/s$\cite{Blakemore}, and with the same parameters as above we find $T_\text{max}^\text{sodium chloride}\approx2\cdot10^{-2}\ K$ as well. That the two crystals, in spite of their very different properties, start to feel the effects of the thermal background at the same temperature is no coincidence; For slightly higher temperatures the black-body peak sits on the frequencies associated to the gravitational wave phonons, and as we lower the temperature the peak is displaced to lower energies. From some critical temperature on, the frequencies of the phonons created by the gravitational wave are in the Boltzmann-suppressed ultraviolet tail of the blackbody spectrum.

The real problem with the above detectors does not come however from the comparison with thermal fluctuation. Indeed, since the relic GW background is expected to be highly anisotropic, we would expect a yearly modulation of the signal which, would allow its detection even in very strong thermal backgrounds \cite{Maggiore:2007}. It is also possible to detect time modulation of the signal of a different source, for instance gravitational radiation produced by cosmic strings. The formation of these strings, which takes place in the very early universe, is accompanied by violent outbursts of gravitational waves. Although the time span of these outbursts is microscopic, of the order of $M^{-1}_{GUT}$ (here $M_{GUT}$ is the symmetry breaking scale of the unified theory in which the strings are embedded), cosmological redshift may stretch this signal in time so that by now it has a duration of years, thus being amenable to detection by the techniques advanced in this paper. However, the energy per unit time scattered in phonons is so small that we are effectively limited by quantum shot noise.  Take for instance sodium chloride. 
For $h\sim10^{-30}$, we have from \eq{edep} $\Gamma=5.7\cdot 10^{-33}\ J/s$ or, in terms of number of phonons per unit time, $2.8\cdot10^{-10}\ s^{-1}$. This means that the mean time between two consecutive phonon scattering events is around $110$ years.  The case for graphene is even worse: We have $\Gamma=8.8\cdot10^{-41}\ J/s$, which corresponds to one phonon scattering event each $7\cdot10^9$ years. The much poorer performance of graphene versus sodium chloride is due to the fact that whereas the former has a higher sound speed and smaller lattice spacing, the other has the advantage of dimension which, for large enough crystals such as the ones being considered here, is decisive.

Quantum shot noise is a fundamental limitation of  any gravitational wave solid state detector, to our knowledge not acknowledged in previous literature. If the signal was weak but continuous, we might be able to extract it using modulation analysis, but the discrete nature of the phonon production process makes this unfeasible.

The problem of quantum shot noise is directly related to the smallness of the amplitude of the GW background at this frequency. 

\section{Exploiting the band structure of the detector}
The above results show that direct detection of the phonons created by a gravitational wave in a solid is not feasible, at least in a straightforward approach. However, the expression \eq{tr} for the transition rate is crucially affected by the band structure properties of the solid. In particular, the last two factors in \eq{tr} can in principle vanish, resulting in a greatly enhanced transition rate. 

The first of them vanishes whenever $k_n'$ is an integer. In other words, the difference between the phonon and GW wave vectors is comparable to a dual lattice basis vector. This corresponds to a gravitational wave of wavelength comparable to the atomic spacing of the crystal. In this regime, the approximations we have taken above when writing the perturbation are not valid. The other possibility for the vanishing of this term is that $k_n'=0$. This corresponds to GW wavelength much larger than the characteristic size of the crystal. As we stated above, in this regime free rather than periodic boundary conditions are appropriate \cite{PhysRev.117.306, grischuk}.  Hence, to be consistent with our approximations, we should consider this term to be always small. 

However, a priori, nothing prevents the last factor in \eq{tr} from blowing up. Recalling the derivation, it can be seen in a straightforward manner that the denominator is nothing but a phase space factor for the final state phonon. Its divergence is a kind of Van Hove singularity \cite{PhysRev.89.1189,Ashcroft}, which shows up as a divergence rather than as a discontinuity in the density of states. This happens because the incoming GW picks a preferred direction which renders the dynamics effectively one-dimensional along this direction.

At a Van Hove singularity, the phase space density of states diverges. Thus, when approaching the singularity along some direction $\b{k}=\b{b}_nk_c$,  the phonon band frequency has a local extremum. Therefore, the generic behaviour of the phonon band near a Van Hove singularity at dimensionless momentum $k_c$ is, to second order in $k_c$ 
\begin{align}\omega_{m}=\frac{\hbar\vert \b{b}_n\vert^2(k-k_c)^2}{2m_*}+\omega_c,\end{align}
a quadratic approximation parametrized by an effective mass along the relevant direction. This leads to a density of states factor
\begin{align}\frac{1}{\vert \nabla\omega_m(\b{k}_{n,m}\cdot\b{b}_n)\vert}=\frac{m_*}{\hbar\vert \b{b}_n\vert^2(k-k_c)}=\sqrt{\frac{m_*}{2\hbar\vert \b{b}_n\vert^2(\omega_g-\omega_c)}}.\end{align}
We have omitted the index $p$ since in our approximation there is only one solution to $\omega_m(\b{k}_{n,m,p}\cdot\b{b}_n)=\omega_g$, namely $k$ (we remind the reader that $p$ is an index running over all such solutions). We thus see that the density of states behaves as $(\omega-\omega_c)^{-1/2}$, this being an integrable singularity. This shows that the divergent behaviour of \eq{tr} will be cured if we consider a non-monochromatic wavepacket. Consider a GW perturbation of the form
\begin{align}h_{ij}=h\epsilon_{ij} \int d\omega_g\ e^{i[\b{k}_g\b{x}-\omega_ g t]} g(\omega_g),\quad\text{with}\quad \int d\omega_g\ \vert g(\omega_g)\vert^2=T.\end{align}
Here, $T$ is some appropriate normalization time which we will take to be the effective time duration of the signal; we will see later on that on general grounds this is bounded by decoherence effects and is not equal to the true duration of the signal. Then, if we denote by $\Delta V_{\omega_g}$ the perturbation to the crystal system created by a monochromatic gravitational wave of frequency $\omega_g$, we are now subjecting the system to a perturbation $\Delta V_g=\int d\omega_g g(\omega_g) \Delta V_{\omega_g}$. Using again Fermi's golden rule, the mean number of phonons produced after the GW has passed will be
\begin{align}N=\hbar^{-2}\sum_{\b{k},m} \left\vert \int d\omega_g g(\omega_g) \int_0^\infty\langle k\vert P_{\omega_g}\ket{0} \right\vert^2=\int d\omega_g \Gamma_{\omega_g} \vert g(\omega_g)\vert^2,\end{align}
where $\Gamma_{\omega_g}$ is given by \eq{tr}. If $\Gamma_{\omega_g}$ were actually independent of $\omega_g$, this would correspond to the usual transition rate $\Gamma\cdot T$. In our present context, however, it is not, and for wavepackets peaked around the Van Hove singularity (i.e. resonant) we can safely take $\Gamma_{\omega_g}\propto \vert\omega_g-\omega_c\vert^{-1/2}$. We then have $N\sim T^{3/2}$ as a function of $T$, the normalization of $g(\omega_g)$. In other words, on-resonance the Van-Hove singularity has the effect of enhancing the phonon transition rate from $T$ to $T^{3/2}$.  To see this, parametrize the $T$ dependence of $g(\omega)$ as $g(\omega)=T f( \omega T)$ for some function $f$, and notice that on-resonance we can make the shift $\omega_g\rightarrow \omega_g+\omega_c$ to have an integral of the form
\begin{align}\int\frac{d\omega_g}{\sqrt{\vert\omega_g\vert}}\vert g(\omega_g)\vert^2=T^{3/2}\int \frac{dx}{\sqrt{x}}\vert f(x)\vert^2.\end{align}

As an example of this enhancement, let us consider a gaussian wavepacket centered at the critical frequency $\omega_c$ and with a width $T$. Let us ignore the contributions of all resonant momenta $\b{k}_{m,n,p}$ except the one sitting at the Van Hove singularity, as well as the order one matrix element taking into account the geometry of the unit cell and GW in \eq{tr}. Considering for the moment that the only relevant dependence comes from the divergent term (this is a good approximation as the other terms only depend weakly on the frequency near $\omega_c$) the expected number of photons behaves as 
\begin{align}N=\frac{T^2}{\pi}\left(\frac{\pi N h^2\bar{m}L_n\vert a_n\vert}{8\hbar}\right)\sqrt{\frac{m_*}{2\hbar\vert\b{b}_n\vert^2}}\int_{\omega_c}^\infty\frac{d\omega_g\ \omega_g^3}{\sqrt{\vert\omega_g-\omega_c\vert}}e^{-T^2(\omega_g-\omega_c)^2}\frac{1}{4\pi^2\left(k_c+\sqrt{\frac{2m_*(\omega_g-\omega_c)}{\hbar \vert\b{b}_n\vert^2}}\right)^2}\end{align}
where we have taken the approximation that $k_n'\approx k_n\ll1$. This means that the phonon wavelength is much smaller than the crystal dimensions (which are within an order of magnitude or two of the GW wavelength), as well as much larger than atomic dimensions. If we assume $T^2$ is large so that the Gaussian is highly peaked around its central value, one can shift the integration variable and evaluate the integral, while evaluating the non-diverging contributions to the integrand at $\omega_c$, which yields (in the $T\rightarrow\infty$ limit)
\begin{align}N=\frac{T^{3/2}}{\pi}\left(\frac{\pi N h^2\omega_g^3\bar{m}L_n\vert a_n\vert}{8\hbar}\right)\sqrt{\frac{m_*}{2\hbar\vert\b{b}_n\vert^2}}\frac{\Gamma\left(\frac54\right)}{2\pi^2k_c^2}.\label{ssss}\end{align}
By comparing the above expression with \eq{gapprox}, we can estimate a value of $c_s$ in \eq{gapprox} which would yield the same number of detected particles as in \eq{ssss}. Thus we arrive at an effective sound speed which parametrizes the efficiency of the detection on-resonance with the Van Hove singularity. One must keep in mind that this way of parametrizing the resonance depends on the specific choice of wavepacket; for reasonable wavepackets the difference will be a factor of order one. We have
\begin{align}c_s^{\text{eff.}}=\frac{\Gamma\left(\frac54\right)}{4\pi^3}\sqrt{\frac{m_*T}{2\hbar}}\frac{(a\omega_g)^2}{k_c^2}.\label{ceff}\end{align}

\section{Detecting gravitational waves}
Let us summarize the requirements for the expression \eq{ceff} to be approximately valid. We need (1) a Van Hove singularity close enough to the origin of the Brillouin zone (so that phonon wavelengths are much larger than the lattice spacing), (2) a gravitational wavepacket on-resonance with the singularity and sharp enough, (3) absence of decoherence so that the phonon amplitude continues to build up for as long as the signal lasts.

This last point is the most challenging from a practical point of view. It is not realistic to expect a $T^{3/2}$ behaviour of the number of scattered particles. Indeed, different scattering events correspond to different phonons and are incoherent to each other, preventing the amplitude \eq{ssss} to be interpreted directly as a transition rate.  In other words, the effective transition rate is $\Gamma=\text{const.}\cdot T_c^{1/2}$, where $T_c$ is some time scale provided by the physics of the system. Since the $T^{3/2}$ law comes from the fact that quantum interference on resonance is constructive, the time scale $T_c$ must be related to some decoherence phenomenon, such as the phonon scattering off some impurities, or reaching the boundary of the crystal and interacting with the detector.

In the real world, interaction with a detector quickly erases quantum correlations within a system. This is so because the detector effectively is an environment for the system, with a macroscopic amount of degrees of freedom, which wash out decoherence. Real modelling of a detector and its environment has been achieved before \cite{PhysRevLett.107.050504,Demers:1995ch}, but we will not need this for our purposes. The essence of decoherence is that, once a system entangles with the environment, the entanglement remains on the macroscopic degrees of freedom, with the probability of it being restored to the system is negligible. Each scattered phonon will interact with the detector on average for a finite amount of time only. Once the phonon is absorbed by the detector, this interaction will be remembered forever by it, preventing a coherent sum of amplitudes from developing. 

It will only be legitimate to use expressions such as \eq{ssss} for times shorter than the decoherence time $T_c$. Since phonon scattering is such a rare event, we can safely address times $T\gg T_c$ by dividing $T$ into steps of duration $T_c$ and consider that they behave independently. Thus, the average total phonon count will scale as $T/T_c$. This is equivalent to substituting $T_c$ in \eq{ceff} and using the resulting transition rate \eq{gapprox} in the usual way. Notice that this means that $T_c$ must be much larger than frequency of the Van Hove resonance, so that the approximation $T\rightarrow\infty$ in deriving \eq{ssss} is justified. 

We may estimate the coherence time of a phonon state as the mean time before an inelastic scattering event happens. Thus, we can take $T_c$ roughly as the mean free path of a phonon in the crystal, divided by the group velocity (effective sound speed at that frequency). We have therefore specified the effective speed of sound entirely in terms of properties of the crystal. Since all the other factors in \eq{ssss} are roughly of the same order of magnitude for most crystals, we can estimate a universal transition rate near a Van Hove singularity of ($L\approx1\ m$, $a\approx10^{-9}\ m$, $m\sim30$ a.m.u), to yield
\begin{align}\Gamma\sim\left(6\cdot10^{34}\ s\cdot m^{-1}\right) h^2 \omega_g c^{\text{eff}}_s.\end{align}
For $h\sim10^{-30}$ and $\omega_g\sim1.8\cdot 10^{11}\ Hz$, this means $\Gamma\sim(c^{\text{eff}}_s/1 m/s)10^{-14}\ s^{-1}$. For $c_s^{\text{eff}}$ of the order of the sound speeds in regular materials we get back to mean times between consecutive scattering phonon events of decades or hundreds of years. But this can be greatly enhanced by the Van Hove singularity. Taking $k_c\sim0.1$,  an effective mass $m_*\sim\frac{\hbar}{2\omega_g}\left(\frac{2\pi}{a}\right)^2$ (this means that the curvature of the band at the singularity is comparable to the band height divided by the length of the Brillouin zone, as many band minima are), coherence time given by the time it takes for a phonon to cross the detector (since at low temperatures boundary effects will be the main source of decoherence),   $T_c\sim10^8\omega_g^{-1}$, we get $c_s^{\text{eff}}\sim10^{6}\ m/s$. Any deviation from the above numbers is expected in any realistic example but this will change the result by no more than a few orders of magnitude.  

With this effective sound speed, $\Gamma\sim10^{-8}\ s^{-1}$ which puts the mean time between phonon detections in a little over three years. Improving the coherence time and taking into account the neglected order 1 factors, it is not unreasonable to expect phonon detection times of months. Detectors of this time resolution and higher open up the possibility of studying a yearly modulation of the signal induced by the motion of the Earth around the Sun. Time modulation analysis allows for the extraction of very small signals over large backgrounds~\cite{Maggiore:2007}.

\subsection{A sapphire monocrystal tower}
\begin{figure}[!hbtp]
\centering
\includegraphics[width=0.65\textwidth]{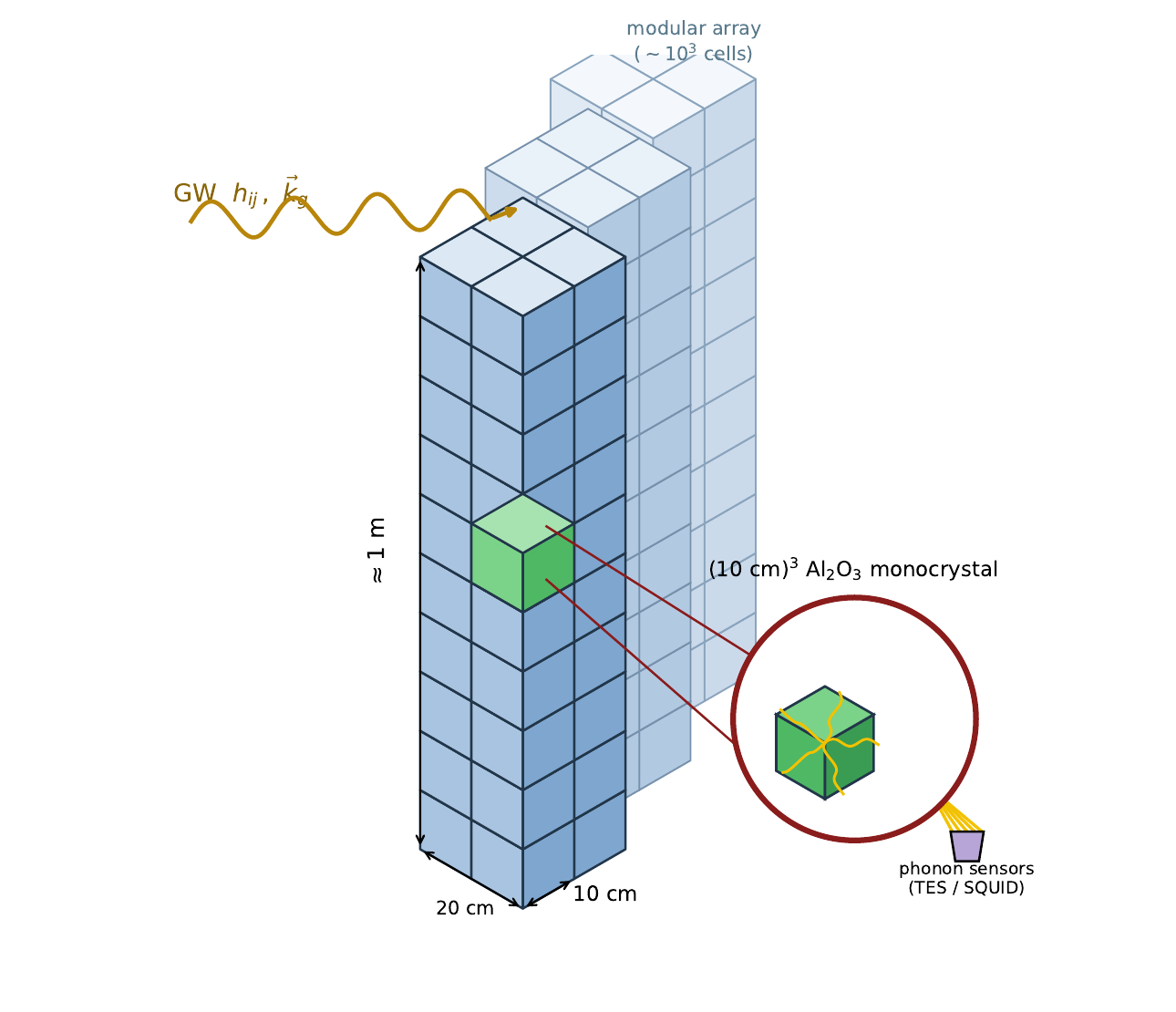}
\caption{Proposed \emph{modular} experimental setup: a tower of $(10\ \mathrm{cm})^3$ sapphire (Al$_2$O$_3$) monocrystal cubes of total height $\approx1$ m, in the spirit of the DAMA/ANAIS detector arrays. Many such towers (lighter, receding) are tiled side by side into a modular array of $\sim10^3$ independent cells forming a full cubic-metre detector. An incoming gravitational wave $h_{ij}$ with wave vector $\b{k}_g$ coherently excites phonons within each cube; the zoom shows a single monocrystal cell read out on its boundary by cryogenic single-phonon sensors (TES/SQUID). The modularity allows each cell to be thermally isolated and read out independently, and provides coincidence/multiplicity-based rejection of cosmic-ray and radioactive backgrounds.}
\label{f2}
\end{figure}

A concrete experimental realization of these ideas can be built by recycling the monocrystal technology already developed for direct dark matter searches such as DAMA/LIBRA and ANAIS \cite{2012RScI...83i1101L}. We envisage an array of ultra-pure sapphire (Al$_2$O$_3$) monocrystal cubes of side $10$ cm, stacked into a tower of total height $\approx1$ m, with each cube instrumented on its boundary by cryogenic single-phonon sensors (transition-edge sensors or SQUID-based calorimeters), as sketched in Fig.~\ref{f2}. Sapphire is chosen for its high sound speed ($c_s\approx10\ \mathrm{km/s}$), small lattice spacing ($a\approx5\cdot10^{-10}\ m$) and the availability of large, dislocation-poor boules. With the parameters quoted above, a single $(10\ \mathrm{cm})^3$ cube already provides a coherent cross section in the sub-millibarn range, and the modular tower geometry allows the effective volume and the boundary detector coverage to be scaled up while keeping each phonon-collection cell small enough that boundary decoherence sets a well-defined coherence time $T_c$. Operating the tower at millikelvin temperatures suppresses the thermal phonon background well below the level estimated above, so that, as discussed above, the sensitivity is ultimately limited by the quantum shot noise of the phonon production process rather than by thermal fluctuations.

Stacking $10\times10\times10$ such cells builds a full cubic-metre detector out of $10^3$ independent $(10\ \mathrm{cm})^3$ monocrystals (the tower of Fig.~\ref{f2} being one representative column). This segmented design is preferable to a single monolithic $1\ \mathrm{m}^3$ block for several reasons. Ultra-pure, dislocation-poor sapphire can be grown routinely at the $10$ cm scale, the size used in the DAMA/LIBRA and ANAIS arrays, while a defect-free metre-sized boule is essentially unattainable; and since the phonon mean free path, which sets the coherent volume entering the $(L/a)^{d+1}$ factor, is limited by impurities and boundaries to at most the cell size, hardly any coherent cross section is lost by segmenting. Each cube can moreover be \emph{thermally isolated} and held individually at millikelvin temperatures, and its small heat capacity ($C\propto V$) is important for single-phonon calorimetry: a transition-edge sensor coupled to a $(10\ \mathrm{cm})^3$ cell has a lower energy threshold and better resolution than one coupled to a whole cubic metre. The $10^3$ cubes also behave as $10^3$ mutually \emph{incoherent}, independent detectors: their contributions add incoherently, $\Gamma_{\rm tot}=\sum_i\Gamma_i\simeq10^3\,\Gamma_{\rm cube}$, and since phonon production is a rare Poisson (shot-noise-limited) process, the statistical significance of the counting experiment grows as $\sqrt{\Gamma_{\rm tot}\,t}$ with the integration time $t$. The modularity further provides background discrimination: a stochastic GW background produces uncorrelated, multiplicity-one hits in single cells, whereas cosmic rays, ambient radioactivity and mechanical noise typically deposit energy in several adjacent cubes at once and can be vetoed by coincidence/anti-coincidence and multiplicity cuts, as in modular dark-matter experiments. Knowing which cell fired also yields coarse position and, statistically, directional information that aids the search for the yearly modulation of the signal. The price is a shorter per-cell coherence time, $T_c\sim L_{\rm cell}/c_s\sim10^{-5}\ \mathrm{s}$ (against $\sim10^{-4}\ \mathrm{s}$ for a metre-sized crystal), which reduces the Van Hove enhancement only mildly, since $c_s^{\rm eff}\propto\sqrt{T_c}$; this $\mathcal{O}(\sqrt{10})$ penalty is outweighed by the gains in crystal purity, phonon readout and background rejection.

\subsection{Choice of detector material: sapphire versus diamond}
The reach of the detector scales with the material combination $\bar{m}\,c_s/a^{d}$ that appears in the cross section $\sigma\propto(2\pi G/c^3)\,\bar{m}\,c_s\,L^{d+1}/a^{d}$, which rewards a high sound speed, a small lattice spacing and light atoms. Among all crystals diamond is the extreme case: it has the highest acoustic velocity of any bulk solid ($c_s\approx18$--$19\ \mathrm{km/s}$, against $\approx10\ \mathrm{km/s}$ for sapphire), the smallest lattice spacing ($a=3.57\ \mathrm{\AA}$), the lightest constituent ($^{12}$C) and the highest Debye temperature known ($\Theta_D\approx2220\ \mathrm{K}$). Combining these, the material factor $\bar{m}\,c_s/a^{3}$ is about three times larger for diamond than for sapphire, so a diamond cell of a given size yields roughly a threefold larger phonon-production cross section. The enormous Debye temperature places the GHz phonons deep in the acoustic band, with a negligible thermal population and a vanishingly small heat capacity ($C\propto(T/\Theta_D)^3$) that is ideal for single-phonon calorimetry; isotopically enriched $^{12}$C diamond, moreover, has the longest phonon mean free path of any solid, which directly benefits the coherence-limited enhancement discussed above. Diamond is also a wide-band-gap semiconductor ($E_g=5.47\ \mathrm{eV}$), so, unlike insulating sapphire, it permits the simultaneous read-out of phonons and ionization and hence an event-by-event discrimination between nuclear and electronic recoils, a powerful handle against radioactive and cosmic-ray backgrounds; cryogenic diamond detectors with sub-$20\ \mathrm{eV}$ energy thresholds and athermal phonon-collection efficiencies exceeding those of sapphire have already been demonstrated in the light dark-matter context \cite{Aggarwal:2025,2012RScI...83i1101L}.

There is, however, a serious practical limitation: large single-crystal diamond does not exist. Synthetic diamond is grown by chemical-vapour deposition or under high pressure and temperature, but monolithic single crystals are restricted to the centimetre scale (commercial plates of $\sim1$--$2\ \mathrm{cm}$, with inch-sized and $50\ \mathrm{mm}$ wafers obtained only by mosaic stitching of several crystals), and the material is orders of magnitude more expensive than sapphire. A $(10\ \mathrm{cm})^3$ diamond monocrystal is simply not available at any price, and even the modular strategy introduced above would require $\sim10^6$ centimetre-sized diamond cells to fill a cubic metre. Sapphire, by contrast, is grown routinely as dislocation-poor boules of tens of centimetres at modest cost. We therefore retain sapphire as the baseline material for a macroscopic, cubic-metre array, while noting that diamond is the ideal medium for a small, high-performance prototype cell in which its superior calorimetric resolution and its phonon-plus-ionization background discrimination could be exploited; scaling coherent diamond targets to macroscopic sizes remains an open materials-science challenge.

\section{Broad-band operation: stochastic backgrounds versus chirps}
The two classes of source discussed in the Introduction --- the stochastic background from the Big Bang and the transient chirp of a light PBH merger --- place very different demands on the detector. We examine here the corresponding detection thresholds and the trade-off between a broad-band and a resonant device, and argue that a solid-state phonon detector is naturally suited to broad-band operation \cite{Aggarwal:2025}.

\subsection{Detection thresholds: stochastic background versus chirp}
A stochastic background (SGWB) is stationary, incoherent and noise-like. In our detector it manifests only as a small, steady excess in the phonon-production rate $\Gamma$. There is no deterministic waveform to match, so the optimal estimator is a counting one \cite{Maggiore:2007}: one accumulates phonon counts over a time $T$ and searches for an excess over the thermal and shot-noise fluctuations, with a significance that grows only as $\mathrm{SNR}\sim\Gamma_s T/\sqrt{(\Gamma_s+\Gamma_{\rm th})T}\propto\sqrt{T}$. The measurement is therefore fundamentally shot-noise limited: each phonon is an independent, isolated count, indistinguishable event by event from a dark count, and the signal can be extracted only statistically and through its sidereal and annual modulation \cite{Maggiore:2007}. This is precisely why the bare background is undetectable --- of order a century between events for sodium chloride --- until the Van Hove resonance raises $\Gamma$.

A PBH merger is the opposite kind of signal: a coherent, deterministic chirp with a known phase evolution, sweeping upward in frequency through the band. Two features relax its detection threshold. First, the known phase permits matched filtering and coherent integration over the cycles spent in band \cite{Maggiore:2007}, so that the amplitude signal-to-noise grows as $\sqrt{N_{\rm cyc}}$ rather than as a raw count --- a coherent gain the stochastic case cannot exploit, although it is ultimately bounded by the decoherence time $T_c$ that limits the usable coherent window. Second, a \emph{nearby} merger is loud: the strain of a local PBH coalescence exceeds the $\sim10^{-30}$ level of the primordial background by many orders of magnitude, so a single passage deposits a large, time-correlated, high-multiplicity burst of phonons --- many cells firing within the chirp duration along a rising time--frequency track. Such an event lies far above threshold and is essentially self-tagging, in sharp contrast to the isolated single counts of the background. The price is that the chirp is transient, sweeping through the GHz band in a short time unless stretched by cosmological redshift, and that loud nearby sources are rare. In short, the background calls for a deep, long-integration, modulation-based search, whereas the chirp calls for a fast, coincidence-tagged, matched-filter search.

\subsection{Broad-band versus resonant detectors}
A high-quality-factor resonance --- whether a Weber bar, a bulk acoustic-wave mode \cite{Goryachev:2014}, or the Van Hove peak discussed above --- accumulates the signal coherently at a single frequency with a build-up factor $\sim Q$, achieving an excellent sensitivity within a narrow bandwidth $\Delta f\sim f/Q$. This is optimal for a monochromatic signal of known frequency, such as a narrow line in the background, but the device is blind elsewhere. A chirp sweeps through the resonance in a time $\Delta t\sim\Delta f/\dot f$ and deposits only the small fraction of its power that falls within $\Delta f$; a single narrow resonator therefore captures almost none of the burst, and covering the sweep would require a comb of resonators tuned across the band, as in the multi-mode bulk acoustic-wave concept \cite{Goryachev:2014}. A broad-band detector instead collects signal wherever the source radiates --- the entire chirp track, or the full width of a broad background --- trading peak spectral depth for frequency coverage. Because the power of both a chirp and a broad background is spread in frequency, the \emph{integrated} signal-to-noise is generally larger for a broad-band device \cite{Aggarwal:2025}. There is thus a tension in our scheme: the Van Hove enhancement is narrow-band, benefiting the background peak at $\omega_c$ but doing little for a chirp.

\subsection{Widening the band of a solid-state detector}
A solid-state phonon detector is, however, broad-band, in contrast to a single-mode resonant bar. The phonon dispersion is continuous, so for any gravitational-wave frequency below the Debye frequency there is a mode to excite, the Bragg-selected $\b{k}=\b{k}_g+\b{b}_n$, and the off-resonance response covers the whole spectrum, from essentially zero up to $\sim\mathrm{THz}$, through the $(L/a)^{d+1}$ continuum \cite{Kahn:2024}. Narrowness arises only when one relies on a single Van Hove singularity for the enhancement, and several strategies can widen or tile the accessible band. One may use the many branches and extrema: a crystal with $\ell$ atoms per cell has $3\ell$ phonon branches, each with several extrema and hence many Van Hove critical frequencies \cite{PhysRev.89.1189,Ashcroft}, so richer unit cells provide a denser comb of resonances. One may also stagger the modular array, giving different cells different lattice spacings, sound speeds, orientations or materials so that their Van Hove frequencies tile the target band, the solid-state analogue of a multi-mode acoustic comb, to which the $\sim10^3$-cell array introduced above is well suited. Crystal orientation can be exploited as well: since the Bragg condition makes the response directional, differently oriented cells resonate at different effective frequencies and along different directions, broadening the band and adding directional information. The upper edge can be raised with a stiffer crystal of higher Debye frequency (diamond, $\omega_D\sim40\ \mathrm{THz}$, against sapphire), a larger sound speed and a smaller lattice spacing pushing the entire phonon spectrum upward. The array can also be operated in a hybrid mode, with some cells run broad-band to follow a chirp by time--frequency (spectrogram) tracking while others are tuned to staggered Van Hove resonances to dig out the narrow background peak. The primordial background is thus best pursued in a deep, narrow, resonance-enhanced, long-integration mode, whereas a PBH chirp is best caught in a broad, fast, coincidence-tagged mode; the modular solid-state array can accommodate both, because the underlying phonon spectrum is continuous and its individual cells can be tuned independently.

\section{Conclusions}
We have studied the use of crystalline solids as detectors of high-frequency gravitational waves in the MHz--GHz band, through the direct production of phonons. Working in a general $d$-dimensional setting, we have computed the phonon-production rate and cross section fully quantum-mechanically, taking into account the band structure of the solid --- the first such calculation, to our knowledge, in contrast to the well-known Weber-bar cross section, which retains only the macroscopic elastic properties of the material. Such a detector is motivated both by the stochastic background from (p)reheating after the Big Bang, peaked at a few GHz, and by the coalescence of light primordial black hole binaries, whose merger frequency $f\simeq4.4\ \mathrm{kHz}\,(M_\odot/M)$ falls in the same band for planetary to sub-planetary masses and would be an unmistakable signature of sub-solar, primordial black holes.

By computing the number of scattered phonons per unit time we have shown that, for a monochromatic signal, the bare rate is far too small to be useful: thermal backgrounds can be suppressed by operating at millikelvin temperatures, but the true obstacle is the quantum shot noise associated with the discreteness of the phonon-production process. For sodium chloride we find a mean time between two consecutive phonon-scattering events of about $110$ years.

The main result of the paper is that this situation improves dramatically through the use of Van Hove singularities in the transition rate. Because Bragg's law renders the gravitational-wave interaction with matter effectively one-dimensional, any band-structure extremum produces an integrable one-dimensional Van Hove singularity; on resonance the phonon-production rate is enhanced by more than an order of magnitude (a factor of a few tens in our worked example, and potentially more for optimized band structures), bringing the mean time between events down to a few years or even months.

On the experimental side we have outlined a concrete, if still preliminary, realization: a modular array of $\sim10^3$ ultra-pure sapphire (Al$_2$O$_3$) monocrystals of $(10\ \mathrm{cm})^3$ each, assembled into a cubic-metre detector in the spirit of the DAMA/ANAIS dark-matter arrays and read out on the boundaries by cryogenic single-phonon sensors. The segmentation is advantageous: it uses crystals of a size that can actually be grown defect-free, lets each cell be thermally isolated and calorimetrically read out with a small heat capacity, and turns the target into $10^3$ independent modules whose Poisson counting statistics and coincidence/multiplicity information provide rejection of cosmic-ray and radioactive backgrounds, as well as sensitivity to the yearly modulation of the signal. Comparing candidate materials, we find that although diamond offers a roughly threefold larger cross section and a semiconductor phonon-plus-ionization channel for background discrimination, the impossibility of growing large single-crystal diamond keeps sapphire as the practical choice for a macroscopic array, with diamond reserved for a small, high-performance prototype cell.

Finally, we have emphasized that the two target signals demand complementary detection strategies. A stochastic background is a stationary, incoherent signal whose detection is a shot-noise-limited counting problem, best pursued in a deep, narrow, resonance-enhanced mode with long integration and modulation analysis; the coalescence of a light PBH binary, by contrast, is a coherent, transient chirp that is loud for nearby sources and deposits a time-correlated, high-multiplicity burst of phonons, best caught in a fast, broad-band, coincidence-tagged mode. A resonant device --- a Weber bar, a bulk acoustic-wave mode, or our own Van Hove peak --- buys sensitivity in a narrow band at the cost of coverage, whereas a solid-state phonon detector is intrinsically broad-band, since its continuous phonon spectrum provides a mode at essentially every frequency below the Debye scale. The modular array is well placed to serve both regimes at once: its cells can be staggered in lattice spacing, sound speed, orientation or material to tile the band with Van Hove resonances, while a subset operated broad-band follows a chirp by time--frequency tracking.

Our estimates rest on several order-of-magnitude approximations, and we have tried to remain conservative; a dedicated study of a specific, realistic crystal and detector model --- building on the mature boundary-phonon technology of dark-matter experiments such as CDMS \cite{2012RScI...83i1101L} --- would be very worthwhile. Our central message is that the gravitational-wave cross section with matter is generically enhanced at singular points of the Brillouin zone, a feature we believe can be exploited in future solid-state gravitational-wave detectors.

\section*{Acknowledgements}

The author thanks Miguel Montero for enlightening discussions on an earlier stage of this project. He also acknowledges support from the Spanish Research Project PID2024-159420NB-C43 [MICINN-FEDER] and the Centro de Excelencia Severo Ochoa Program CEX2020-001007-S at IFT.

\appendix
\section{GW-crystal interaction in general TT gauge}\label{appendixgauge}
The results of the main text have been developed in a particular coordinate system attached to the center of mass of the crystal, and making use of the geodesic deviation equation. This is justified only if the GW wavelength is much larger than the dimensions of the crystal. In the main text we have extended the result to wavelengths comparable to the size of the crystal, just by introducing the spatial modulation in the GW profile. This appendix provides a justification of this procedure, by working in a general TT gauge, not tied to the CM frame of the crystal. This is precisely what is done in other kinds of GW detectors, such as interferometric detectors \cite{Finn:2008np}, in which the GW wavelength is comparable to the size of the detector.

We would like to find the form of the GW-atom interaction in a general TT coordinate system. First, we realize that, even though the GW may be comparable in size to the crystal, we are always in the regime where $\lambda_g\gg a$, with $a$ being the typical interatomic spacing in the crystal. Since an atom only has significant interactions with its close neighbours, we can always sit in the CM frame associated to the equilibrium position of one atom, and consider the interaction with its close neighbours to be again
\begin{align}V_{\b{R},l}&=\sum_{\b{R}',l'}u^j_{l'}(\b{R}')D_{ij}^{ll'}(\b{R}-\b{R}'+l-l')u^j_{l}(\b{R})\nonumber\\&=\sum_{\b{R}',l'}[x^j_{l'}(\b{R}')-R'-c_{l'}]^jD_{ij}^{ll'}(\b{R}-\b{R}'+l-l')[x^j_{l}(\b{R})-R-c_{l}].\end{align}
Here, we are dropping the direct interaction terms between the GW and the atoms; since the atoms under consideration are only a few lattice parameters apart, the gravitational forces between them are negligible. But we can now transform back to our original TT gauge using the standard transformation \cite{MTW}
\begin{align}x^j=(\delta_i^j+\frac{1}{2}h_j^i)x^i_{TT}.\end{align} 
Doing so, the potential in TT coordinates takes (to first order) the form 
\begin{align}V&=\sum_{\b{R},l}V_{\b{R},l}\nonumber\\&\approx \frac{1}{2}\sum_{\b{R},l,\b{R}',l}u^j_{l'}(\b{R}')D_{ij}^{ll'}(\b{R}-\b{R}'+l-l')u^j_{l}(\b{R})+h_j^\nu (R'+c_{l'})^jD_{\nu\mu}^{ll'}(\b{R}-\b{R}'+l-l') u^\mu_l\b{R},\end{align}
where we have dropped the $TT$ subindex since no ambiguity arises here.
But now this procedure may be repeated for each unit cell in the crystal, and in each step we determine the interactions of the atoms of that particular lattice with its neighbours. Adding up all the contributions we find the GW-crystal interaction in the TT gauge,
\begin{align}V&=\sum_{\b{R},l}V_{\b{R},l}\nonumber\\&\approx \frac{1}{2}\sum_{\b{R},l,\b{R}',l}u^j_{l'}(\b{R}')D_{ij}^{ll'}(\b{R}-\b{R}'+l-l')u^j_{l}(\b{R})+h_j^\nu (R'+c_{l'})^jD_{\nu\mu}^{ll'}(\b{R}-\b{R}'+l-l') u^\mu_l\b{R},\end{align}
where we have dropped terms both quadratic in $h$ and $u$, since these will give negligible contributions.

From the above we see that in the TT the interaction is what we would have if there was no gravitational wave, plus an interaction 
\begin{align}\Delta V'= \frac{1}{2}\sum_{\b{R},l,\b{R}',l}h_j^\nu (R'+c_{l'})^jD_{\nu\mu}^{ll'}(\b{R}-\b{R}'+l-l') u^\mu_l\b{R}.\end{align}
Using the equations of motion, $P'$ may be rewritten as
\begin{align} \Delta V'&=\frac{1}{2}\sum_{\b{R}',l}h_j^\nu (R'+c_{l'})^j\left[\sum_{\b{R},l}D_{\nu\mu}^{ll'}(\b{R}-\b{R}'+l-l') u^\mu_l\b{R}\right]\nonumber\\&=-\frac{1}{2}\sum_{\b{R}',l}m_{l'}h_j^\nu (R'+c_{l'})^j\left[\sum_{\vec{k},m}\omega_m(\b{k})^2 \alpha_m(\b{k}) \epsilon^i_{l',m}(\b{k})e^{i[\b{k}\b{R}'-\omega_m(\b{k})t]}+\text{conj.}\right].\label{ppr}.\end{align}
Finally, let us assume a plane, monochromatic GW, $h_{ij}=he^{-i[\b{k}_g\b{x}-\omega_gt]}\epsilon_{ij}$. In terms of annihilation operators, the interaction is
\begin{align}\Delta V'=-\frac{h\epsilon_{ij}e^{i\omega_g t}}{2}\sum_{\b{R},l,\b{k},m}m_l\ \omega_m(\b{k})^2\sqrt{\frac{\hbar}{2N\bar{m}\omega_m(\b{k})}}[R+c_l]^j\left( a_m(\b{k}) \epsilon^i_{l,m}(\b{k})e^{-i\b{k}_g[\b{c}_l+\b{u}_l]} e^{i[\b{k}-\b{k}_g]\b{R}}+\text{H.c.}\right).\label{P2}\end{align}

At first sight it seems that \eq{P2} cannot be right, because it does not reduce to \eq{P} in the $\b{k}_g\rightarrow0$ limit. There is no contradiction here, as we are comparing perturbations in different gauges, and these will be in general gauge-dependent. In passing from one gauge to another we must realize that the wavefunction of the system will undergo a gauge transformation. In fact, the relationship between the CM frame and the TT gauge is very much the same as the relationship between the dipolar and the Coulomb gauge in standard electrodynamics \cite{Bal,PhysRevA.40.3764}. In this case, it is well-known that naive transition amplitudes in the Coulomb gauge only coincide with those of the dipolar gauge on-resonance; for off-resonance amplitudes one also has to take into account the gauge transformation of the wavefunction. However, in the text we only compute on-resonance amplitudes (since we only concern ourselves with phonon production at late times compared with the typical scattering length of phonons, i.e. we use Fermi's golden rule), and therefore can effectively substitute $\omega_m(\b{k})$ by $\omega_g$ in \eq{P2}; in this case, we do recover \eq{P}, the only difference being that the phase factor changes to take into account the (now non-negligible) GW momentum, as advertised in the main text. But then, the factors $e^{-i\b{k}_g[\b{c}_l+\b{u}_l]}$ may be dropped since the GW wavelength is much larger than the lattice parameter.

\bibliographystyle{apsrev4-1}
\bibliography{references}
\end{document}